\documentclass[useAMS,usenatbib]{mn2e}

\usepackage{graphicx}

\title[\emph{K2} photometry]{Precise time-series
  photometry for the Kepler-2.0 mission}
\author[S. Aigrain et al.]{S. Aigrain$^{1}$\thanks{E-mail:
suzanne.aigrain@astro.ox.ac.uk}, S. T. Hodgkin$^{2}$,
M. J. Irwin$^{2}$, J. R. Lewis$^{2}$, and S. J. Roberts$^{3}$\\
$^{1}$Department of Physics,
University of Oxford, Keble Road, Oxford, OX1 3RH, UK\\
$^{2}$Institute of Astronomy, University of Cambridge, Madingley Rise,
Cambridge, CB3 0HA, UK\\
$^{3}$Department of Engineering Science,
University of Oxford, Parks Road, Oxford, OX1 3PJ, UK}

\begin{document}

\date{Accepted \ldots Received \ldots; in original form \ldots}

\pagerange{\pageref{firstpage}--\pageref{lastpage}} \pubyear{2014}

\maketitle

\label{firstpage}

\begin{abstract}
  The recently approved NASA \emph{K2} mission has the potential to
  multiply by an order of magnitude the number of short-period
  transiting planets found by \emph{Kepler} around bright and low-mass
  stars, and to revolutionise our understanding of stellar variability
  in open clusters. However, the data processing is made more
  challenging by the reduced pointing accuracy of the satellite, which
  has only two functioning reaction wheels. We present a new method to
  extract precise light curves from \emph{K2} data, combining
  list-driven, soft-edged aperture photometry with a star-by-star
  correction of systematic effects associated with the drift in the
  roll-angle of the satellite about its boresight. The systematics
  are modelled simultaneously with the stars' intrinsic variability
  using a semi-parametric Gaussian process model. We test this method
  on a week of data collected during an engineering test in January
  2014, perform checks to verify that our method does not alter
  intrinsic variability signals, and compute the precision as a
  function of magnitude on long-cadence (30-min) and planetary transit
  (2.5-hour) timescales. In both cases, we reach photometric
  precisions close to the precision
  reached during the nominal Kepler mission for stars fainter than
  $12^{\rm th}$ magnitude, and between 40 and 80 parts per million for
  brighter stars. These results confirm the bright prospects for
  planet detection and characterisation, asteroseismology and stellar
  variability studies with K2. Finally, we perform a basic transit
  search on the light curves, detecting 2 bona fide transit-like
  events, 7 detached eclipsing binaries and 13 classical variables.
\end{abstract}


\section{Introduction}

The NASA \emph{Kepler} mission, launched in 2009, collected wide-field
photometric observations of over 150\,000 stars located in a
single, 110-sq.\,deg.\ field of view (FOV) continuously over 4
years. The unprecedented precision, baseline and homogeneous nature of
this dataset has revolutionised the fields of exoplanet and stellar
astrophysics. The analysis of the first 22 months of data alone has
led to the detection of thousands of planet candidates, ranging from
super-Jupiter to sub-Earth size, including some located in the
habitable zone of their host stars
\citep{bor+11a,bor+11b,bat+13,bur+14}, and the analysis of the full
4-year dataset is still underway. However, the failure of the second
of four reaction wheels in 2013 forced the nominal mission, which
relied on exquisite pointing control, to end prematurely. After a
period of tests to evaluate the pointing performance and photometric
precision achievable with only two reaction wheels, the satellite has
now embarked upon a new mission, called \emph{Kepler}-2.0, or
\emph{K2}. Over the next 2--3 years, \emph{K2} will observe 4 fields
per year for $\sim 80$\,days each. As the two remaining reaction
wheels control the pitch and yaw, the thrusters must be used to
alter the roll angle of the spacecraft about its boresight. The K2
fields are thus located in or near the orbital plane of the spacecraft
(which is approximately in the Ecliptic plane) so as to minimize the
torque induced by solar pressure about the boresight of the
telescope. The thrusters are fired every few hours to correct
the roll angle to its nominal value.

The science goals, observing strategy and preliminary performance of
the \emph{K2} mission are described in \citet[][hereafter
H14]{how+14}. Briefly, the top-level science goals include:
\begin{itemize}
\item the search for planetary transits around bright $R \le 12$
  Sun-like stars and around low-mass (late K and
  early M) stars (\emph{K2} will observe roughly as many of these in
  each pointing as \emph{Kepler} did in its entire lifetime);
\item the search for transiting planets around, and other forms of
  variability in, tens of thousands of stars in a wide range of
  Galactic environments (compared to the single direction observed by
  \emph{Kepler});
\item observations of nearby open clusters and star forming regions,
  including asteroseismology, rotation and activity studies, and
  searches for eclipsing binaries and transiting planets;
\item observations of a wide range of extragalactic sources including
  supernovae.
\end{itemize}

Aside from the different pointing strategy, \emph{K2} will be operated
in a similar way to \emph{Kepler}. Small `postage stamps' will be
extracted from the raw images around each target of interest, and
co-added on board to result in a cadence of 29.4\,min (long-cadence)
for most stars, before downlinking to the ground. A small subset of
the targets will be observed in short (1\,min) cadence mode. Owing to
the reduced pointing accuracy, the postage stamps collected have to be
larger than during the nominal \emph{Kepler} mission, so the total
number of targets observed per field is much smaller (10--20\,000 per
run, compared to $\sim$160\,000 for \emph{Kepler}). Beyond this there
are two important differences in the way targets are selected for
observation and the way the data is processed. The targets for each
run are proposed by the community at large and allocated on the basis
a ranking of their scientific value. Where a large number of targets
are proposed in a relatively small area of sky (e.g. in an open
cluster), multiple postage stamps can be collected together to form a
so-called `superstamp'. Additionally, for the first year of operations
at least, the data will be released to the community in the form of
target pixel files (TPFs) rather than light curves. TPFs are sequences
of postage stamp images calibrated for known, pixel-level effects
using the first stages of the \emph{Kepler} data processing pipeline.

The rest of the pipeline used during the nominal \emph{Kepler} mission
to extract light curves and correct them for instrumental systematics,
as well as to search for and validate planetary transits, will need
some modifications before they can be applied to \emph{K2}. In the
mean time, the community is being encouraged to develop and use
alternative light curve extraction and calibration methods, applying
them to the TPFs which will be released approximately 6 weeks after
the end of each \emph{K2} observing run.  The present paper presents
one such method, a purpose-written pipeline designed to extract light
curves from \emph{K2} photometry and correct them for systematic
effects associated with the pointing drift of the satellite.

To develop and test our method, we used the so-called two wheel
engineering test dataset\footnote{\tt
  http://archive.stsci.edu/missions/k2/tpf\_eng/.}, which was recently
released by the \emph{K2} science office for this purpose. This
dataset consists of 2079 long-cadence and 18 short-cadence TPFs, each
containing $50 \times 50$ pixel postage stamps, collected over a
period of 9 days, corresponding to 440 long-cadence and 13200
short-cadence observations. The TPFs were created from the data
downloaded from the spacecraft by the \emph{Kepler} pipeline
\citep{jen+10}. The present paper focusses on the
long-cadence observations (our method can be extended to short-cadence
data, but this is deferred to a later paper).  While most TPFs were
centred on a specific star, two `superstamps' were created using
blocks of contiguous TPFs, one near the centre of the FOV and one near
the edge. The first $\sim 2.5$\,days of data were collected in coarse
pointing mode (i.e. using external star trackers to control the
spacecraft pointing) and the remainder in fine pointing mode (i.e.\
using the fine guidance sensors mounted on the focal plane to control
the pointing).

The paper is structured as follows: the methods used to extract the
photometry and calibrate the systematics are presented in
Sections~\ref{sec:phot} and \ref{sec:roll_corr}, respectively. The
results are discussed in Section~\ref{sec:res}, where we evaluate the
photometric precision of the light curves as a function of magnitude,
and discuss the behaviour for variable stars. In
Section~\ref{sec:disc} we compare the performance of our pipeline to
other published methods for extracting photometry for \emph{K2}, and
carry out a transit search on stars brighter than $14^{\rm th}$
magnitude. Finally,
Section~\ref{sec:concl} summarises our results and their implications for
the \emph{K2} key science goals, and outlines areas for future work.

\section{Light curve extraction}
\label{sec:phot}

\subsection{List-driven aperture photometry}

The approach we adopt to extract photometry from the \emph{K2} TPFs is
based on experience gained in the context of several ground-based
wide-field time-series photometric surveys, namely: the Monitor
project \citep{aig+07}, the UKIRT WFCAM Transit Survey (WTS,
\citealt{nef+12}), and the
VISTA variables in the \emph{V{\'i}a L{\'a}ctea} survey (VVV,
\citealt{sai+12}). In all of these, we obtained the most satisfactory
results in terms of photometric precision using \emph{list-driven
  aperture photometry}, which consists of:
\begin{enumerate}
\item running source detection software to identify the sources in
  each image or frame, and measuring their centroid positions. This
  results in a catalog of sources for each image.
\item deriving a precise astrometric solution by comparing the
  catalogs extracted from each image to a master catalog, which can be
  constructed from a master image (e.g. a stack of a subset of the images taken
in the best conditions, if extracting faint sources is important) or a
publicly available catalog. 
\item performing soft-edged aperture photometry on each frame for each source in the master
  catalog, by placing a circular aperture at the location of the source
  on that image, as computed from its sky coordinates (more details on
  the aperture photometry are given below).
\end{enumerate}
We attribute the success of this method so far to its relative
simplicity, and to the fact that the global astrometric solution used
is much more precise than the individual centroid position
measurements for each star, particularly for fainter stars (typical
uncertainties in the astrometric solution for Nyquist-sampled images
are of order 0.1 pixels).

The basic aperture function used is a circular top hat with no
tapering. Flux from pixels completely enclosed within the aperture
boundary are given full weight ($w_i = 1$).  The soft-edging refers to
those pixels intersected by the boundary.  For these the
background-corrected flux $f_i$ is split pro-rata according to the
fractional area $w_i$ of the pixel enclosed within the boundary.  For
an isolated image, this is simply equivalent to defining the total
flux as $\sum_i f_i \times w_i$.  For overlapping or blended images,
we treat the circular top-hat as a point-spread function (PSF) and do
simultaneous PSF fitting using the same pro-rata split to define the
individual model PSFs across boundaries.

Although circular apertures/PSFs are not optimal for all images, they are
close to optimal for cases where the Poisson noise from the objects is the
dominant noise contributor. They also have the advantage of minimising the
impact of systematics caused by uncertainties, or any mismatch, in the
exact shape of the PSF.  For example, conventional weighted PSF fitting
introduces  a magnitude-dependent systematic offset between total flux and
PSF-derived flux in such cases which is absent in our aperture-based
approach.

\subsection{Implementation for \emph{K2}}

To apply this method to \emph{K2} data, we must first construct full frame
images (FFIs) from the \emph{K2} TPFs. To do this, we start from a \emph{Kepler} FFI
(taken during the nominal mission). \emph{Kepler} FFIs are multi-extension
FITS files with 4 extensions for each of the 21 science modules
(each module consists of 2 $1048 \times 1048$\,pixel CCDs, each of
which is divided into 2 output channels). For each \emph{K2}
long-cadence observation, a new copy of the original \emph{Kepler} FFI is
created, and all the image arrays are set to zero. For each TPF, the
postage stamp image corresponding to this observation is then inserted
at the appropriate location in the relevant extension (using the
meta-data contained in the TPF headers). For each extension, we also
used the {\sc python} module {\sc k2fov}\footnote{\tt
  http://keplerscience.arc.nasa.gov/K2/ToolsK2FOV.shtml.} to compute
the sky coordinates of the 4 corners of each output channel, and from
these we obtained a first estimate the {\sc crval} and {\sc cdx\_x}
header keywords describing the World Coordinate System (WCS) solution,
which were inserted in the extension header. We also generated a
single binary mask with the same format as the FFIs, which is 1 where
there is data and 0 where there is none. We used this mask as a 
confidence map, and note that it could be refined further to flag bad pixels and
account for effects such as vignetting over the field of view, if
needed.

We then performed source extraction, astrometric solution and
list-driven aperture photometry on the FFIs using the {\sc imcore},
{\sc wcsfit} and {\sc imcore-list} routines from the {\sc casutools}
package\footnote{\tt
  http://casu.ast.cam.ac.uk/surveys-projects/software-release}
\citep{irw+04}, using our binary mask as a confidence map. Note that
the confidence map could also be used to flag bad pixels and account
for effects such as vignetting over the field of view, if needed. The
source extraction software normally makes a local estimate of the
background at each point in the image using the median of a $64 \times
64$-pixel window, which is larger than the \emph{K2} postage stamps;
we therefore proceed using a single, global background estimate for
each extension, but bearing in mind a more careful background
correction may be needed at a later stage. For example, this could be
done by fitting a simple surface to the background variation over each
detector or output channel (the latter might be needed if differences
in bias correction between channels corresponding to the same
detector).

The astrometric solution was obtained relative to
the 2MASS all-sky point-source catalog, using the catalog extracted
from the first image as the master catalog. This resulted in a total
of 13\,977 sources, of which 13\,633 were detected (at the
4-$\sigma$-level on $>50\%$ of the images. The typical root mean
square (RMS) of the astrometric solution was 0.4\arcsec, or
approximately 0.1 pixel, which is satisfactory given the under-sampled
nature of the \emph{Kepler} point-spread function (PSF).

The pipeline evaluates the flux of each source on each image by
placing a circular aperture centred at exactly the same \emph{sky}
position on every frame, derived from the global astrometric
solution. In our experience, this approach gives systematically better
results than using the centroid or best-fit position of each source as
measured on each frame to locate the aperture. Our apertures are
soft-edged, in the sense that pixels straddling the edge of the
aperture contribute partially to the flux, as opposed to the
pixellated masks used by the \emph{Kepler} pipeline. Finally, the
pipeline automatically deblends sources with overlapping apertures,
assigning the appropriate fraction of the flux in the overlap pixels
to each aperture by simultaneously fitting a `top hat' function to all
overlapping sources in a blend.

\paragraph{Aperture radii:} We computed fluxes using 6 different
aperture radii: 1.5, 3, $3\sqrt{2}$, 6, $6\sqrt{2}$ and 12
pixels. In general, larger apertures are more suitable for brighter
sources, as the flux in the wings of the PSF remains significantly
above the background out to larger radii. The optimal aperture radius
to use for a given star also depends on its individual neighbourhood:
if a source has a close neighbour, a very small aperture which
excludes as much of the neighbour's flux as possible \emph{or} a very
large one which encompasses both sources might give better results
than an aperture of intermediate size. However, different apertures
can be more or less sensitive to certain systematic effects, and yield
better performance depending on the main timescales of interest. We
therefore process all the aperture fluxes in parallel for every star.
Throughout the paper, we use the 3-pixel apertures, unless stated
otherwise. This radius corresponds to approximately twice the
full-width at half-maximum (FWHM) of the PSF, which in our experience
gives good results for most sources across a wide range of magnitudes,
although one should bear in mind that it may not be optimal at either
end of the magnitude range. We return to the question of how to select the
aperture radius more in Section~\ref{sec:aprad}.

\subsection{Catalog cross-matching and zero-point calibration}

In order to be able to identify known variables among our sources, and
to be able to place the objects on a standard magnitude scale, we cross-matched the
list of objects for which we extracted light curves (our master
catalog) with the target list supplied with the test dataset
(hereafter, the input target list). This was done by locating, for
each object in the latter, the closest positional match in the
former. For 56 of the 1951 targets in the input target list, the
closest match in the master catalog was located further than 8\arcsec
away, i.e.\ approximately two \emph{Kepler} pixels. Some of these
could be faint objects which were not detected by our source detection
software, but a number are relatively bright. These could be high
proper motion objects where the the nominal position in the input
target list does not match the actual position at the time of the
observations. 

We then calibrated the flux measurements obtained using each aperture radius
onto the \emph{Kepler} magnitude scale by fitting, for all the matches closer
than 8\arcsec, a relation of the form:
\[
K_{\rm p} = {\rm zp}_X - 2.5 \log ({\rm flux}_X)
\]
where $K_{\rm p}$ is the \emph{Kepler} magnitude and the subscript $X$ refers
to the different aperture radii. The
zero points obtained in this way were all between 25 and 25.3 (for
fluxes expressed in e$^-$/s). For faint objects with nearby
neighbours, the magnitude is strongly dependent on the aperture size
used. Therefore, for the remainder of this paper, we adopt a single
magnitude for each object, based on the 3-pixel aperture fluxes.

We checked if there were any trends in the flux versus magnitude
relation depending on the distance of the source from the satellite
boresight. Such a trend might be expected due to the fact that the
images of stars located near the corners of the FOV are significantly
elongated, which could cause some of the flux to fall outside the
aperture, but we found no evidence for it.

\begin{figure}
  \centering
  \includegraphics[width=\linewidth]{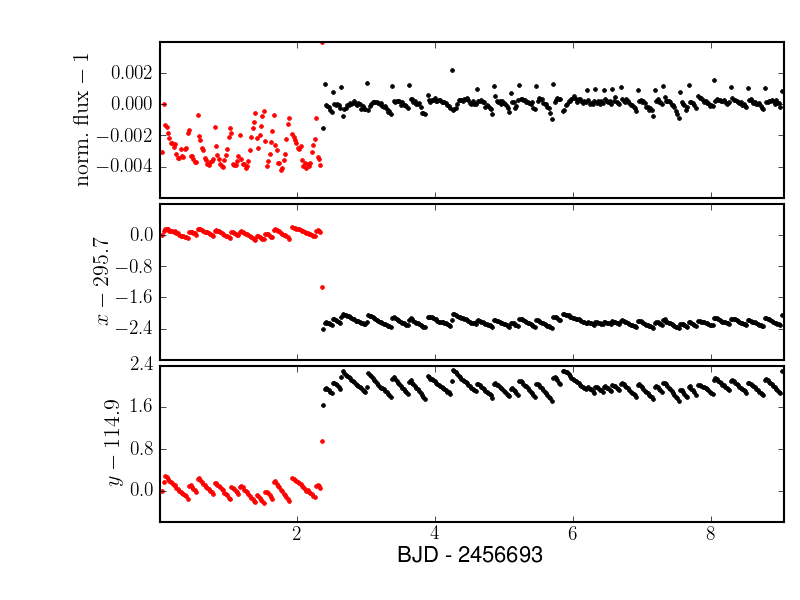}
  \caption{Example raw light curve and centroid coordinates (EPIC
    no.\ 60021515, cool dwarf, \emph{Kepler} magnitude $K_{\rm p} = 11.3$,
    located in module 9, output channel 1). Data taken
    before fine-pointing was achieved are shown in red.}
  \label{fig:lcraw1}
\end{figure}

\begin{figure}
  \centering
  \includegraphics[width=\linewidth]{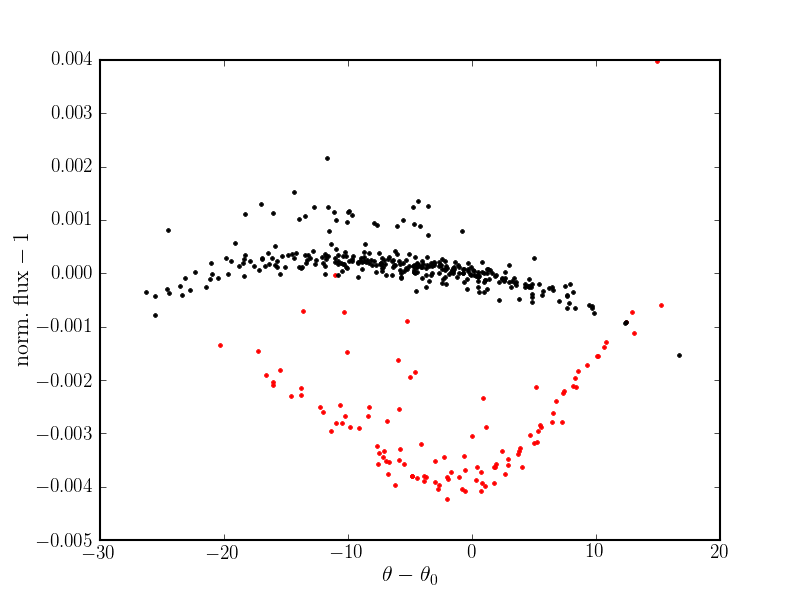}
  \caption{Same as Figure~\protect\ref{fig:lcraw1}, but now showing
    flux versus the roll angle $\theta$ of the spacecraft. Here
    $\theta$ is the mean of the roll angle estimates derived from the
    astrometric solution for each channel, relative to a reference
    frame (which is simply the first frame in the dataset). Note the
    tight but nonlinear relationship between flux and roll angle,
    which differs before and after fine-pointing was achieved, as the
    star is located on a different set of pixels.}
  \label{fig:lcraw2}
\end{figure}

\subsection{Raw light curves}

Figure~\ref{fig:lcraw1} shows an example light curve resulting from
the procedure described in the previous section. Most light curves
show significant variations of instrumental origin, which are clearly
correlated with the object's position on the detector. These
variations, which presumably are due to a combination of intra-pixel
and inter-pixel (flat-field) sensitivity variations, must be corrected
before the light curves become astrophysically useful.  

The pitch and yaw of the spacecraft are fairly stable, as they are
controlled by the two remaining functioning reaction wheels. However,
the roll about the boresight displays a gradual drift due to the Sun's
radiation pressure, and is reset approximately every 6 hours using the
spacecraft's thrusters. This causes the star to wander over the
detector by a significant fraction of a pixel, and leads to systematic
variations in the measured flux. Indeed, as shown in
Figure~\ref{fig:lcraw2}, the systematic flux variations are tightly
correlated with the roll angle (they also show a correlation with the
star's individual $x$- and $y$-position, but less tight). Note that we
estimated the roll angle in each frame for each output channel from
the {\sc crval} parameters of the astrometric solution, and then
adopted the median of the individual channel values as our global roll
angle estimate.

\section{Systematics correction}
\label{sec:roll_corr}

We experimented with a number of established methods for modelling
these roll-dependent systematic effects, starting with simple linear
decorrelation of the flux against roll angle, or the $x$- and $y$-
coordinates of the star's centroid as measured on each image. We also
tried using the well-known SysRem algorithm \citep{tam+05}, where the
systematic trends are iteratively identified as the first principal
component of the matrix constructed from all the light curves, and
each trend is removed by linearly decorrelating the flux measurement
for each star against it. The systematics correction in the PDC-MAP
pipeline used during the nominal \emph{Kepler} mission \citep{smi+12}
is a variant on this approach, where a Bayesian framework is used to
ensure that the coefficients relating each star's light curve to each
trend are similar for stars of similar magnitude and position on the
detector, as might be expected is their physical origins are
similar. However, no PDC-MAP data are publicly available at the time
of writing for this dataset.  Finally, we also tested the
`Astrophysically Robust Correction' (ARC) method proposed by
\citet{rob+13}, once more in the context of the nominal \emph{Kepler}
mission. This is again a variant on SysRem-like approaches, where
additional measures are used at the trend detection and removal stages
to minimize the risk of removing variability of astrophysical
origin. However, none of these approaches gave satisfactory results:
in some light curves the systematics were only partially corrected,
while in others systematics were actually introduced by the
correction.

\subsection{Gaussian process model}

All of the methods described above implicitly assume that the
relationship between the trends and the stellar fluxes is linear, and
this is not generally the case for \emph{K2} data, as can be seen in
Figure~\ref{fig:lcraw2}. In fact, while flux is often correlated with
roll angle, the amplitude and shape of the relationship between the two
varies from star to star, even for stars of similar brightness located
close to each other on the detector. This is because inter- and intra-pixel
sensitivity variations (flat-field) are the dominant source of
systematics for \emph{K2}. The exact manner in which the roll angle
variations impact the star's measured flux depends on the response of
the specific set of pixels on which each star falls. 

Systematics removal approaches based on linear basis modelling can be
used to remove non-linear effects, so long as the behaviour can be
described by a linear combination of such models and the driving
systematic variables can be identified. However, we do not know
\emph{a priori} whether these conditions apply for the present
dataset. We therefore opted to model the systematics using a Gaussian
Process (GP) with roll angle as the input variable. This approach
assumes that the systematics are related to the roll angle variations,
but makes no assumptions about the functional form of that dependence:
instead it marginalises over all possible functions, subject to
certain smoothness constraints. Each star is modelled individually, so
the functional form of the systematics is allowed to vary from star to
star. This also enables us to model the intrinsic variability of each
star at the same time as we fit for the systematics model, in an
effort to minimize the extent to which the systematics correction
interferes with the real astrophysical signals.

We omit here a detailed explanation of GP regression, as many
are available elsewhere. For example, the interested reader will find
a thorough, textbook-level introduction to GPs in \citet{rw06}, or a
more succinct introduction in the context of modelling systematics in
high-precision stellar photometry in \citet{gib+12}.
For our purposes, a GP can be thought of as setting up a probability
distribution over functions, whose properties are constrained by
parametrising the covariance between pairs of observations. 
Specifically, we postulate that pairs of flux
measurements taken at similar values of the roll angle (i.e. when the
star fell at approximately the same position on the detector) should
be strongly correlated, while pairs of flux measurements taken at very
different values of the roll angle should be essentially
uncorrelated. This is expressed mathematically by the covariance
function, or kernel:
\[
k_\theta(\theta_i,\theta_j) = A^2_\theta \exp
\left[ - \frac{|\theta_i-\theta_j|^2}{2L^2_\theta}\right]
\]
This form of kernel, known as a squared exponential (SE), is one of
the most commonly used, and gives rise to smooth variations which a
characteristic length scale $L_\theta$ and amplitude $A_\theta$.
The likelihood of observing a sequence of $N$ flux measurements
$\mathbf{y} \equiv \{y_1,y_2,\ldots,y_N\} $ is then a multivariate Gaussian:
\[
\mathbf{y} \sim \mathcal{N} (
\mathbf{0};\mathbf{K}),
\]
where the mean of the distribution has been set to zero, and the
individual elements of the covariance matrix $\mathbf{K}$ are given by
the kernel for the corresponding inputs. Note that we work with light
curves that have been normalised by subtracting their median and
dividing them by a robust estimate of their scatter, estimated as
$\sigma_{\rm MAD} = 1.48\,{\rm MAD}$, where ${\rm MAD}$ is the median
of the absolute deviation from the median.

In practice, the measured flux does not depend on roll angle alone: the light
curve also contains intrinsic variations of astrophysical origin,
which depend on time, and observational (white) noise. We therefore adopt a
composite kernel function of the form
\[
k(\{t_i,\theta_i\},\{t_j,\theta_j\}) = k_t(t_i,t_j) +
k_\theta(\theta_i,\theta_j) + \sigma^2 \delta_{ij},
\]
where
\[
k_t(t_i,t_j) = A^2_t \exp
\left[ - \frac{|t_i-t_j|^2}{2L^2_t}\right]
\]
represents the intrinsic variations, $\sigma^2$ is the white noise
variance and $\delta$ is the Kronecker delta function. The additive
form of the adopted kernel enables us to separate the different
contributions from roll angle, time, and white noise. 
The use of an SE kernel with a single length scale and amplitude to
model the intrinsic variability of each star, which can be quite complex, is an
oversimplification, but it is sufficient for the purpose at hand,
which is merely to help disentangle between intrinsic and instrumental
variations.

\subsection{Implementation for this dataset}

As shown on Figures~\ref{fig:lcraw1} and \ref{fig:lcraw2}, the data collected in coarse
pointing mode present larger roll angle systematics, with a different
relation between flux and roll angle to that observed for the data
collected in fine sampling mode. For the remainder of this paper, we
therefore focus only on the data collected in fine pointing
mode. However, we note that the same procedure could be applied
separately to the data collected in coarse pointing mode, though the
time-dependent component of the GP would not be well-constrained
given the short duration of that data segment.

\subsection{Initial guesses, parameter optimisation and outlier
  rejection}

\begin{figure}
  \centering
  \includegraphics[width=\linewidth]{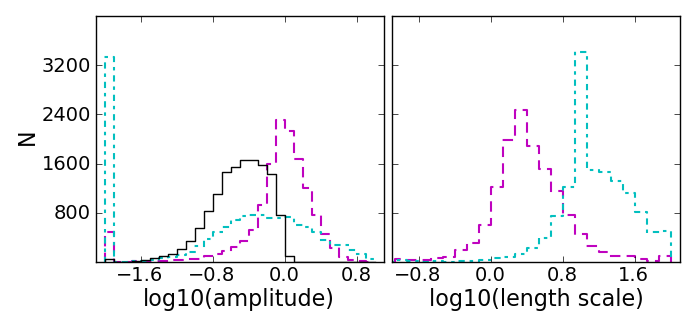}
  \caption{Histogram of the best-fit GP amplitude and length scale
    parameters (left and right panels, respectively). The amplitudes
    are relative to the scatter of the original (uncorrected) light
    curve. The dashed magenta, dash-dot cyan and solid black lines
    correspond to the time, roll angle and white noise components, respectively.}
  \label{fig:GP_par}
\end{figure}

For each light curve, we start with a prior for the
covariance parameters $\{A_t,\,L_t,\,A_\theta,\,L_\theta,\,\sigma\}$
which, after visual inspection of several tens of randomly selected
light curves, was set to $\{0.5,2\,{\rm days},\,0.2,\,10\,{\rm
  arcsec},\,0.2\}$. These values give an adequate description of most
light curves, though they are not suitable for the final
correction. We then evaluate the mean and variance of the predictive
distribution of the GP conditioned on the data, initially without
adjusting the covariance parameters. Any data points lying further
than $3\,\sigma$ from the predictive mean are flagged as likely
outliers. This step is necessary as most light curves contain a number
of outliers can otherwise significantly affect the determination of
the best-fit covariance parameters. We have not established the
precise origin of these outliers, but they tend to occur near the
times where the roll angle of the satellite was reset using the
thrusters. Such a manoeuvre could cause certain parts of the detector
to heat up, and/or smear the stellar flux over a larger area, both of
which could lead to an anomalous flux measurement for the
corresponding cadence.

We then optimize the likelihood with respect to the covariance
parameters using a standard local optimisation algorithm within
pre-set bounds (specifically, the {\sc scipy} implementation of the
{\sc L-BGFS-B} algorithm by \citealt{byr+95}), and ignoring the
outliers flagged at the previous step. The amplitudes were constrained
to be $>0.01$ and the length scales were constrained to lie in the
range 0.1--100. Fig.~\ref{fig:GP_par} shows histograms of the
resulting best-fit GP parameters. Except for stars which display
significant variability on timescales similar to the roll variations
(which are discussed in more detail in Section~\ref{sec:variable}), the
best-fit parameters are insensitive to the initial guesses. We then
repeat the outlier flagging procedure, this time using a more
stringent threshold of $2.5\,\sigma$, before evaluating the mean of
the predictive distribution of the GP a final time. To isolate the
component of the flux variations caused by the roll variations, the
predictive distribution is evaluated using the actual values of the
roll angle, but setting the time to a fixed value (e.g.\ the middle of
the time interval spanned by the observations). This is then
subtracted from the original flux measurement, leaving a light curve
corrected for roll-dependent systematics.

\subsection{Computational cost and subsampling}

Fitting for the covariance parameters requires multiple evaluations of
the likelihood. For a GP of this type, the computational cost of each
likelihood evaluation is $\mathcal{O}(N^3)$. To speed up the
calculation, we subsample the data before fitting for the covariance
parameters. For the test dataset, we used every $4^{\rm th}$ data
point, which enables each light curve to be processed in 0.5--1\,s on
a $1.6$\,GHz Intel processor without any noticeable effect on the
results. More severe subsampling combined with simultaneous processing
on multiple cores may be needed for the longer science campaigns.

\subsection{Corrected light curves}

\begin{figure*}
  \centering
  \includegraphics[width=\linewidth]{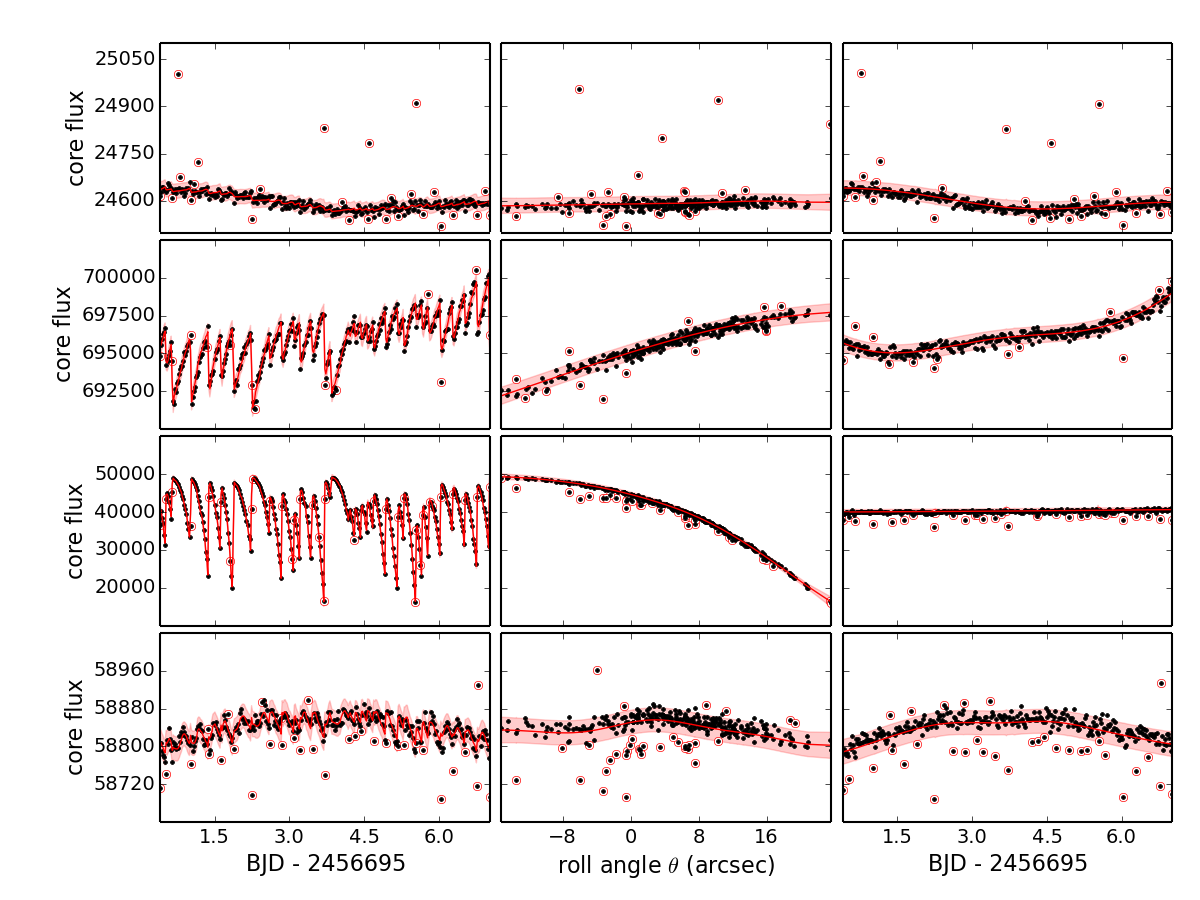}
  \caption{Four examples illustrating the correction of
    roll-dependent systematics using a GP. In the left column, black dots
    show the input light curve and the red curve shows the mean of the
    predictive distribution of the full model (roll-dependent
    systematics $+$ intrinsic variations). Points falling
    $>2.5\,\sigma$ away from this line (i.e.\ outside the pink shaded
    area) are flagged as outliers and marked by the red circles. The
    middle column shows flux as a function of roll
    angle, after subtracting the time-dependent component of the model
    (i.e.\ showing only the systematics). Finally, the right column shows
    the corrected light curve after
    subtracting the roll-dependent component of the model.}
  \label{fig:lcex1}
\end{figure*}
 
Figure~\ref{fig:lcex1} shows four example light curves, before and
after correction of the roll-angle systematics. These examples we
drawn from a random selection, within which we chose four cases
spanning a wide range of magnitudes and locations within the FOV,
where the systematics were visible by eye. They include a clearly
systematics-dominated case (third row), cases where the systematics
have a similar amplitude to the intrinsic (time-dependent) variability
(second and fourth row) and a case where both are similar in amplitude
to the white noise (top row). We visually examined hundreds of similar
examples, and while the systematics are not always strong enough to be
noticeable, we found no instance where the correction seemed to
introduce spurious effects.

\section{Results}
\label{sec:res}

\begin{figure*}
  \centering
  \includegraphics[width=\linewidth]{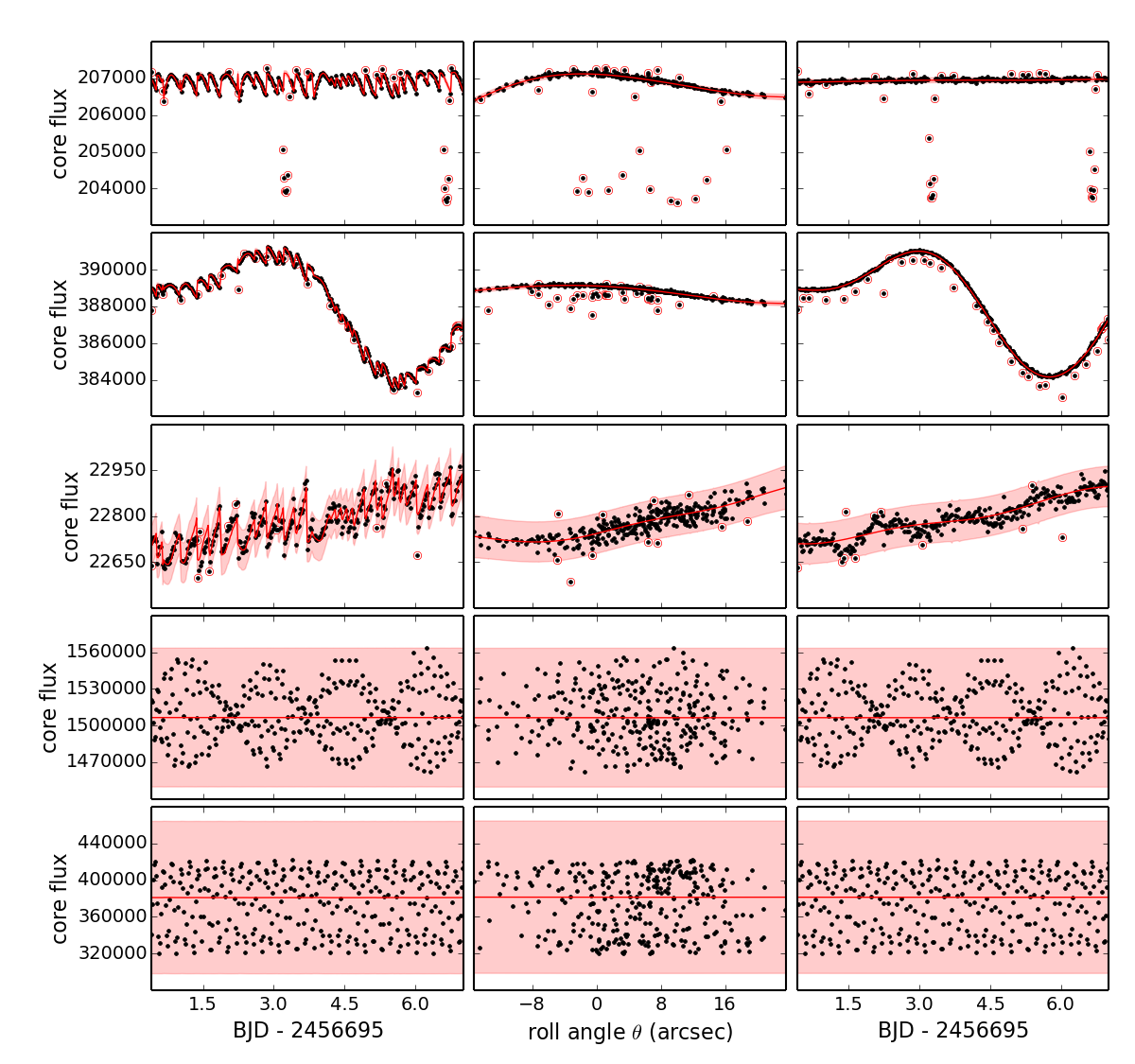}
  \caption{As Figure~\ref{fig:lcex1}, but now showing selected
    variable stars. From top to bottom: known transiting exoplanet host
    star WASP-28 (EPIX 60017806, $K_{\rm p}=12.0$) , a cool dwarf showing slow
    variations most likely caused by star spots (EPIC 60022118, $K{\rm
      p}=11.2$), a star displaying low-level stochastic variability (EPIC
     60017902, $K_{\rm p}=14.4$), a classical pulsator (EPIC 60018082, $K{\rm
      p}=9.7$), and a contact eclipsing binary (EPIC 60017809, $K{\rm
      p}=11.5$). See text for a more detailed discussion.}
  \label{fig:lcex2}
\end{figure*}

\subsection{Effect on transits and variable stars}
\label{sec:variable}

Figure~\ref{fig:lcex2} illustrates the behaviour of our roll-angle
correction for a few selected variable stars. 

\begin{figure}
  \centering
  \includegraphics[width=\linewidth]{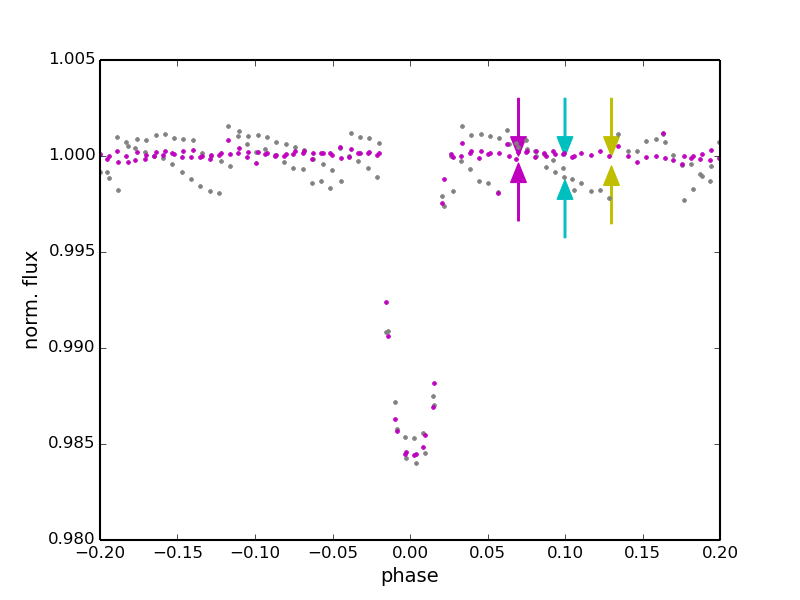}
  \caption{Phase-folded light curve of WASP-28 (using the ephemeris of
    \protect\citealt{and+14}). Grey and magenta dots show the raw and roll-corrected
    data, respectively (long cadence, in both cases). The magenta
    arrows indicate the depth of a transit that would have been
    detectable in this light curve with a signal-to-noise ratio of 8
    (assuming 3 transits lasting 3 hours each). For comparison, the cyan
    and yellow arrows indicate the depths of transits caused by a
    Neptune-sized planet in front of a Sun-like star, and an
    Earth-sized planet in front of a $0.4\,M_\odot$ M-dwarf,
    respectively.}
  \label{fig:wasp28}
\end{figure}

The top panel shows the known transiting planet host star WASP-28. Our
outlier detection procedure flags the in-transit points, and enables
the roll-angle correction to proceed using the remaining
data. Importantly, while the in-transit points are flagged as
outliers, the roll-angle correction is still applied to the full light
curve, including the transits. The phase-folded light curve is shown
in Figure~\ref{fig:wasp28}, illustrating the drastic noise reduction
brought about by the correction of the roll-angle systematics. 

The second panel shows a cool star displaying smooth variations on
timescales of a few days (probably caused by star spots) as well as
significant roll-angle systematics. The intrinsic variability is
reproduced without difficulty by the time-dependent component of our
GP model, enabling the correction of the systematics to proceed
unimpeded by the intrinsic variability. This example illustrates the
fact that our method is suitable for stellar rotation and activity studies.

The third panel shows a star
displaying lower amplitude, more stochastic variability. In this case,
only the long-term component of the variability is modelled, and the
stochastic behaviour that is not well explained by the roll-angle
variation is incorporated in the white noise term. Nonetheless, the
roll-angle systematics are corrected successfully, illustrating the
fact that our method is suitable for studies of stochastic behaviour
such as accretion related variability in young stars.

The last two panels show stars displaying periodic variability on
timescales very similar to the roll angle variations (a pulsating star
and a contact eclipsing binary, respectively). In those cases, our
model does not reproduce the variability, and neither does it identify
(and hence correct) any roll-angle systematics. The amplitude of the
time and roll-angle terms becomes negligible, while the white noise
term becomes anomalously large. These objects are easy to identify (in
Figure~\ref{fig:GP_par}, they fall in the lowest time and roll-angle
amplitude bins \emph{and} in the highest white noise amplitude
bin). Furthermore, they are not pathological, as the light curve is
unchanged by the correction, so no spurious effects are introduced.
Nonetheless, anyone interested in studying this specific type of
object at very high precision would need to develop a different
approach to identify and remove roll-angle systematics.  One way of
doing this might be to model the periodic variability explicitly
(using a sum of sinusoidal terms), \emph{simultaneously} with the
roll-angle systematics (using a GP). Alternatively, phase dispersion
minimisation (PDM, \citealt{ste78}) could be used to remove the periodic
variability without explicitly modelling it, before correcting the for
the roll angle systematics, and then adding the periodic component
back onto the residuals.

\subsection{Photometric precision}

\begin{figure*}
  \centering
  \includegraphics[width=\linewidth]{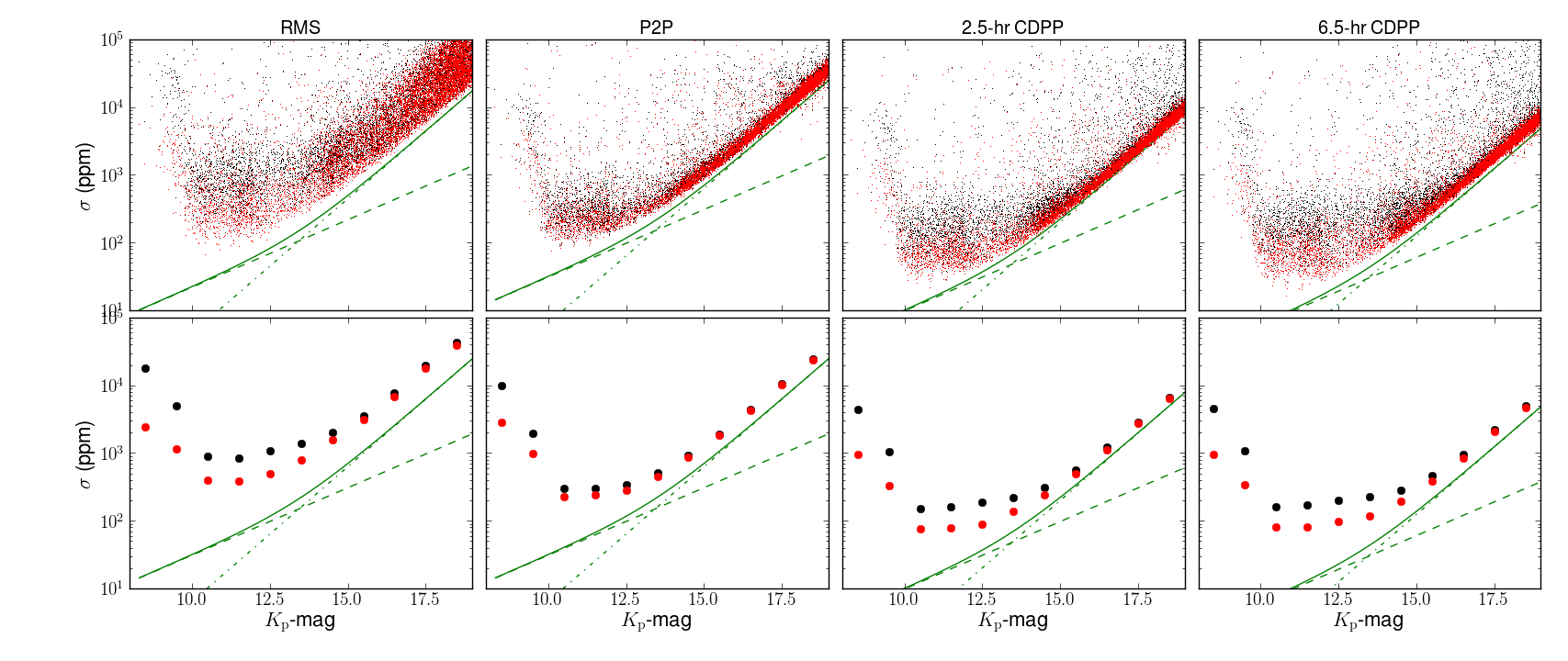}
  \caption{Relative photometric precision (in parts per million) as a
    function of magnitude, before (black) and after (red) correction
    of the roll-angle systematics, using 3-pixel apertures. From left
    to right: full light curve scatter, point-to-point scatter,
    2.5-hour and 6.5-hour quasi-CDPP (see text for details). The top panel
    shows all objects, while the bottom shows the median values in
    1-mag bins. The green solid, dashed and dash-dot lines show the
    theoretical total, source, and background $+$ readout noise,
    respectively.}
  \label{fig:rms_mag}
\end{figure*}

\begin{table*}
  \centering
  \begin{tabular}{rr|rrrr|rrrr}
    \hline
    Mag.\ range & No.\ stars & \multicolumn{4}{|c|}{3-pixel apertures} & \multicolumn{4}{|c|}{$3\sqrt{2}$-pixel apertures} \\
    & & RMS & P2P & 2.5-hr CDPP & 6.5-hr CDPP & RMS & P2P & 2.5-hr CDPP & 6.5-hr CDPP \\
    \hline
    9.0--10.0 &158 & 1158 & 976 & 331 & 343 & 556 & 249 & 105 & 150 \\      
    10.0--11.0 & 402 & 400 & 227 & 75 & 81 & 340 & 182 & 75 & 113 \\       
    11.0--12.0 & 541 & 383 & 238 & 78 & 80 & 359 & 201 & 74 & 86 \\        
    12.0--13.0 & 366 & 500 & 286 & 90 & 99 & 545 & 279 & 81 & 68 \\        
    13.0--14.0 & 439 & 801 & 452 & 136 & 117 & 1130 & 516 & 143 & 111 \\     
    14.0--15.0 & 998 & 1580 & 867 & 242 & 193 & 2400 & 1110 & 302 & 231 \\    
    15.0--16.0 & 1014 & 3093 & 1841 & 498 & 383 & 4922 & 2544 & 685 & 514 \\    
    16.0--17.0 & 1403 & 6896 & 4230 & 1121 & 840 & 11401 & 6030 & 1610 & 1232 \\ 
    17.0--18.0 & 2302 & 17944 & 10301 & 2737 & 2059 & 28547 & 13759 & 3712 & 2821 \\
    18.0--19.0 & 3578 & 38752 & 23673 & 6304 & 4713 & 57747 & 29125 & 7770 & 5856 \\ \hline
  \end{tabular}
  \caption{Median photometric precision as a function of \emph{Kepler}
    magnitude, in parts per million (ppm), for two aperture sizes.}
  \label{tab:precres}
\end{table*}

We evaluated the photometric precision of the light curves using four
different metrics. The first is simply the light curve scatter (once
again, estimated as $\sigma_{\rm MAD}$) -- this is referred to as
RMS. The second is the point-to-point scatter, estimated in the same
way but from the first difference between consecutive points in the
light curve. This estimate, referred to as P2P, includes only the
high-frequency component of the noise, multiplied by a factor
$\sqrt{2}$. The third and fourth are intended to approximate the
combined differential photometric precision (CDPP) used throughout the
\emph{Kepler} mission to quantify the photometric precision on
planetary transit time-scales. However, we cannot reproduce the actual
CDPP values without access to the transit search component of the
\emph{Kepler} pipeline, since the CDPP is essentially the depth of a
transit of a given duration that would give a signal-to-noise ratio of
1 \citep{chr+12}. Instead, we defined a `quasi-CDPP', which we
evaluate as the median of the standard deviation of the light curve
evaluated in a moving window of a given duration. This procedure is
identical to that adopted by \citet[][hereafter VJ14]{vj14}, the only
other study published to date on the photometric precision attainable
with this \emph{K2}. We chose to adopt the same approach to enable us
to compare our results directly with theirs, but note that our
`quasi-CDPP' estimates are equivalent to the actual CDPP only if the
noise is white on the timescale under consideration\footnote{For the
  sake of conciseness, figure labels and table column headings still
  use the shorthand `CDPP').} As the top-level goal of the nominal
\emph{Kepler} mission was the detection of habitable planets, the CDPP
was typically measured on 6.5-hour timescales. The limited duration of
\emph{K2} campaigns (up to 85 days) means that it will be most
sensitive to relatively short-period planets with transit durations of
2--3\,hours, so we report quasi-CDPP values for both 2.5-hour (5
cadences) and 6.5-hour (13 cadences).

\begin{figure}
  \centering
  \includegraphics[width=\linewidth]{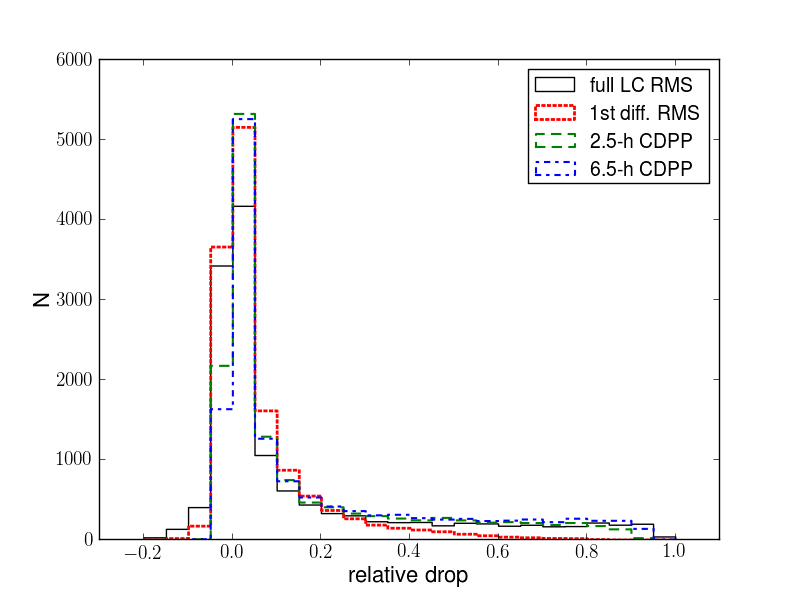} \hfill
  \caption{Histogram of the relative reduction in scatter as measured
    by the full light curve RMS (solid black line), the RMS of the
    first difference (dotted cyan line)
    and the 6.5-hour quasi-CDPP (dashed magenta line).}
  \label{fig:scatter_red}
\end{figure}

Figure~\ref{fig:rms_mag} shows these four different estimates of
the photometric precision as a function of magnitude before and after
correction of the roll-angle systematics, while
Figure~\ref{fig:scatter_red} shows histograms of the relative
reduction in these different precision estimates between the original
and corrected light curves (for 3-pixel apertures in both
cases). Additionally, Table~\ref{tab:precres} gives the median
precision estimates in 1-magnitude bins.

On Fig.~\ref{fig:rms_mag} we also show the estimated source and
background photon noise and readout noise as a function of magnitude,
as well as the total theoretical noise estimate obtained by adding
them in quadrature. In the left panel, the two contributions were
estimated as the square root of the median flux for each source, and
the median background flux integrated over the aperture, respectively,
after converting the fluxes from e$^-$/s to e$^-$ by multiplying by
the duration of individual on-board exposures (6.02\,s) times the
number of exposures per cadence (270 for long-cadence data). For the
P2P and quasi-CDPP plots, the noise estimates were scaled by a factor
$\sqrt{2}$, $1/\sqrt{5}$ and $1/\sqrt{13}$, respectively.

A number of features are worth remarking on. For all metrics used, the
lower envelope of the scatter versus magnitude relation corresponds
approximately to the photon noise level down to magnitude $K_{\rm p}
\sim 15$, and is within a factor 2 of it down to $K_{\rm p} \sim
12$. For brighter stars, the scatter versus magnitude plots flatten
off, indicating that an additional source of systematics is
present. One possible additional noise source is detector response
non-linearity, though using larger apertures for these stars may also
help alleviate the problem. The effects of saturation become visible
for $K_{\rm p} < 10$.

When isolating the scatter on a given timescale, as in the last 3
columns of Fig.~\ref{fig:rms_mag}, the vast majority of stars follow a
tight relationship between precision (after correction) and
magnitude. This indicates that most of the scatter above this level
observed in the raw light curves can be explained by roll-angle
systematics. By contrast, the full light curve scatter (RMS) results
from the combined effects of intrinsic variability on all timescales,
residual instrumental effects and white noise. As a result, a wide
range of RMS values are observed for a given magnitude even after
correction.

The amount by which the correction reduces the full light curve
scatter and the quasi-CDPP varies widely from star to star, ranging
from $<10\%$ for most stars, to $>90\%$ for a few hundred cases. These
tend to be bright stars displaying little intrinsic variability, or
variability on long (days) timescales only, where the systematics were
initially dominant. The point-to-point scatter is reduced by a smaller
amount, on average, because the dominant timescale of the roll angle
variations is significantly longer than the interval between
consecutive observations.

\subsection{Injection tests}
\label{sec:inj}

\begin{figure}
  \centering
  \includegraphics[width=\linewidth]{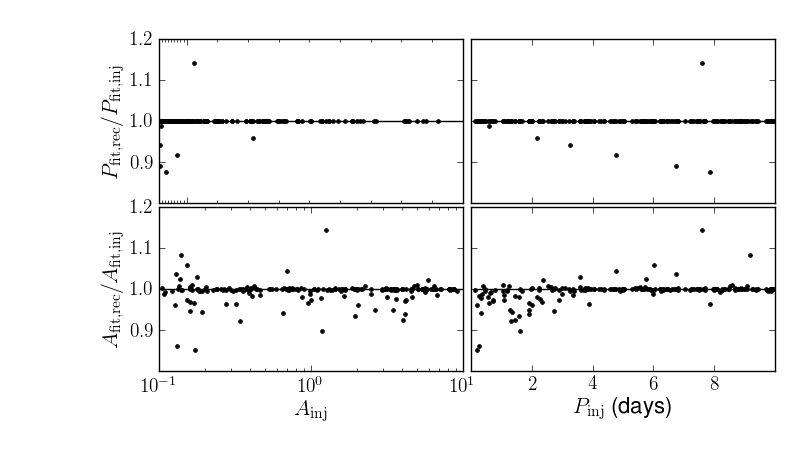}
  \caption{Results of the injection tests (see Section~\protect\ref{sec:inj} for details).}
  \label{fig:inj}
\end{figure}

We have already shown qualitatively in Figure~\ref{fig:lcex2} that our
systematics correction preserves most forms of astrophysical
variability. Here we test this in a more quantitative manner by
injecting sinusoidal signals into the light curves before correcting
them.  This test was performed on 200 randomly selected light curves
with $10 \le K_{\rm p} \le 14$, with the default (3-pixel) aperture
only. The injected signals had periods drawn from a uniform
distribution between 0.1 to 10 days, amplitudes drawn from a log
uniform distribution between 0.1 and 10 times the standard deviation
of the original light curve, and phases drawn from a uniform
distribution between 0 and $2 \pi$.

To assess how well the injected signals were preserved, we performed a
least-squares fit of a sinusoid to injected and recovered signals,
where the recovered signal is defined as the difference between the
corrected light curves with and without injected signal. (Note that
the amplitude and period of this fit can differ from the injected
values, even for the original time-series, owing to the limited
duration of the dataset.) The results of the test are shown in
Figure~\ref{fig:inj}, which shows the ratio of fitted periods and
amplitudes in the original and recovered signals, as a function of the
injected amplitude and period. The periods are recovered perfectly
(i.e.\ up to the frequency resolution of the data) in 99.5\% of the
cases. In the remaining 0.5\% of cases, the discrepancy (which
nonetheless never exceeds 15\%) arises from the fact that the signals
were injected into light curves already containing significant
variability. As the time-dependent component of our model has a single
characteristic time-scale, light curves containing time-variability on
multiple, very different timescales can be problematic. The amplitude
is recovered to within 1\% in 78\% of the cases. The amplitude
recovery performance is significantly worse for periods $<3$\,days,
which are more difficult to disentangle from the roll-angle
variations, but the amplitude is always recovered to better than
15\%. This means that there should be no difficulty in
\emph{identifying} such signals in the detrending light curves, but
joint modelling of systematics and astrophysical signals is advisable
to obtain accurate amplitudes.

\subsection{Aperture selection}
\label{sec:aprad}

As previously mentioned, the best aperture radius to use is expected
to decrease with increasing magnitude. Indeed, when we compute the
aperture giving the smallest scatter -- using any of the 4 metrics
described above -- for each object individually, the general trend is
for smaller apertures for the fainter objects, and larger apertures
for the brighter objects. The exact aperture size selected depends on
the timescale considered: the P2P, 2.5-hr and 6.5-hr quasi-CDPP statistics
favour increasingly larger apertures at a given magnitude. A plausible
explanation for this is that the P2P statistic is primarily sensitive
to white (background and readout) noise, which is minimized by using
smaller apertures, while the quasi-CDPP, particularly on longer timescales,
is more sensitive to any residual pointing-related systematics, which
are minimised by using slightly larger apertures. This is also visible
in Table~\ref{tab:precres}, which lists the median precision metrics
in each magnitude bin for two aperture sizes.

\begin{figure}
  \centering
  \includegraphics[width=\linewidth]{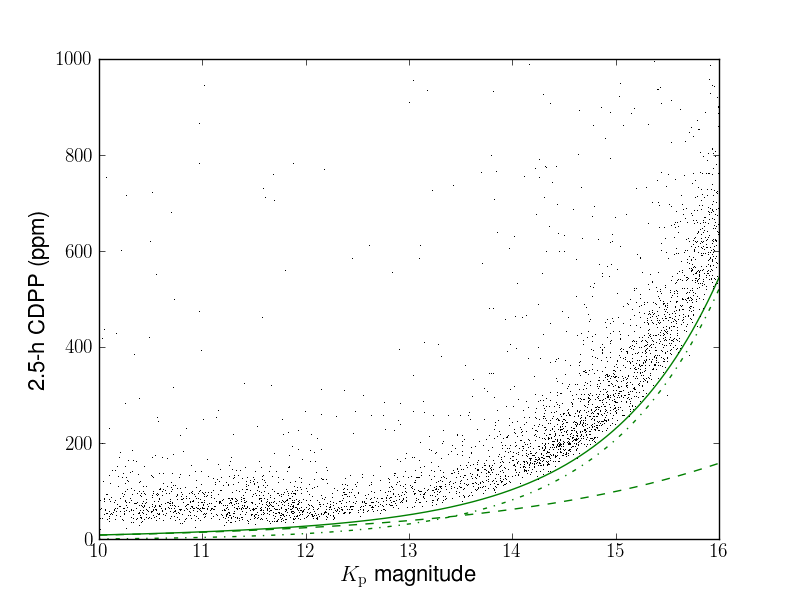} \hfill
  \caption{2.5-hour quasi-CDPP of the light curves corrected for roll-angle
    systematics, using a magnitude-dependent aperture radius
    selection. The green solid, dashed and dash-dot lines show the theoretical
    total, source, and background $+$ readout noise, respectively.}
  \label{fig:cdpp}
\end{figure}

A blind application of the object-by-object aperture selection can
have undesirable consequences, however, particularly the faint end,
where very large apertures would be selected for a significant
fraction of the objects, effectively minimising their scatter by
diluting them with the flux of a nearby neighbour. Taking this into
account, we ultimately implemented a simplified magnitude aperture
selection based on the 2.5-hr quasi-CDPP, but adopting the most
commonly selected aperture size in each magnitude bin for all stars in
that bin, and enforcing a monotonic increase of aperture size with
magnitude. This results in 12-pixel apertures down to $K_{\rm p}=9$,
$3\sqrt{2}$-pixel apertures down to $K_{\rm p}=13$, 3-pixel apertures
down to $K_{\rm p}=16$, and 1.5-pixel apertures beyond
that. Figure~\ref{fig:cdpp} shows the resulting 2.5-hr quasi-CDPP as a
function of magnitude for $10<K_{\rm p}<16$. This choice of apertures
yields performance close to the theoretical noise limit throughout
that wavelength range. The use of a small number of fixed aperture
sizes introduces `jumps' in precision at the magnitudes where the
aperture radius changes, which are visible in
Figure~\ref{fig:cdpp}. These might be avoided by using an aperture
radius that varies smoothly with magnitude. However this would require
a signficiant re-write of our pipeline, and is therefore deferred to
future work.

\section{Discussion}
\label{sec:disc}

\subsection{Comparison to other methods}

At the time of writing, photometric precision estimates for \emph{K2}
are available from H14 and (\citealt{vj14}, hereafter VJ14). 

H14 performed analysed data collected during engineering tests in
October 2013 and January 2014 (the latter are the test dataset used in
the present paper). Using the publicly available {\sc PyKe} package
developped by the \emph{Kepler} Guest Observer office, including
simple aperture photometry using circular or elliptical apertures and
background estimates based on the median of nearby pixels, they
achieved a precision on 6-hour timescales of $\sim 80$\,ppm for
12$^{\rm th}$ magnitude G dwarfs, compared to $\sim 20$\,ppm for the
same stars during the nominal \emph{Kepler} mission. Overall, they
found that the precision over 6-hour timescale was within a factor of 4
of the precision achieved during the nominal \emph{Kepler}
mission. 

VJ14 analysed the January 2014 test dataset only, using a
purpose written pipeline, which performs simple aperture photometry
followed by a correction for systematic effects due to the satellite's
pointing variations on a star-by-star basis. They achieved median
photometric precisions (as measured by the 6.5-hour quasi-CDPP) within a
factor of 2 of the nominal \emph{Kepler} performance, a factor of 2
improvement over the preliminary results presented in H14. 

Our results are broadly similar to those of VJ14; we obtain slightly
better performance for fainter stars, where our results approach the
theoretical noise limit, and slightly worse for brighter stars, for
which the \emph{lower envelope} (rather than the median) of our quasi-CDPP
distribution is within a factor of 2 of the median \emph{Kepler}
performance (and of the photon-noise limit). This last point may be
due to the fact that VJ14 analysed bright stars using very large
apertures, whose position was determined by the centroid of each star
on each frame. We did try increasing our aperture radius up to 12
pixels, but without significant improvement on the resulting
precision.

In principle, our method has several advantages over that of VJ14. It
provides light curves for every target on silicon; it should be less
sensitive to errors in the measurement of the position of individual
stars; the model used for systematics correction is probabilistic,
principled, and enables a rigorous propagation of the uncertainties
associated with the systematic correction. On the other hand, more
work is needed to understand the slightly better performance obtained
by VJ14 for bright stars.

\subsection{Effect of image distortion across the FOV}

\begin{figure*}
  \centering
  \includegraphics[width=0.8\linewidth]{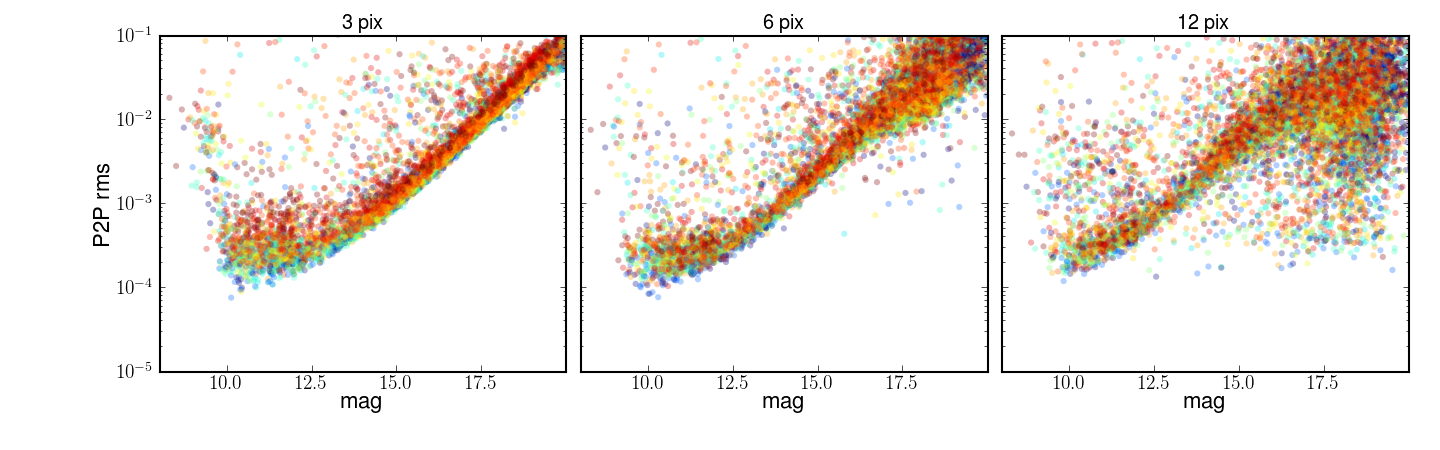}
  \includegraphics[width=0.8\linewidth]{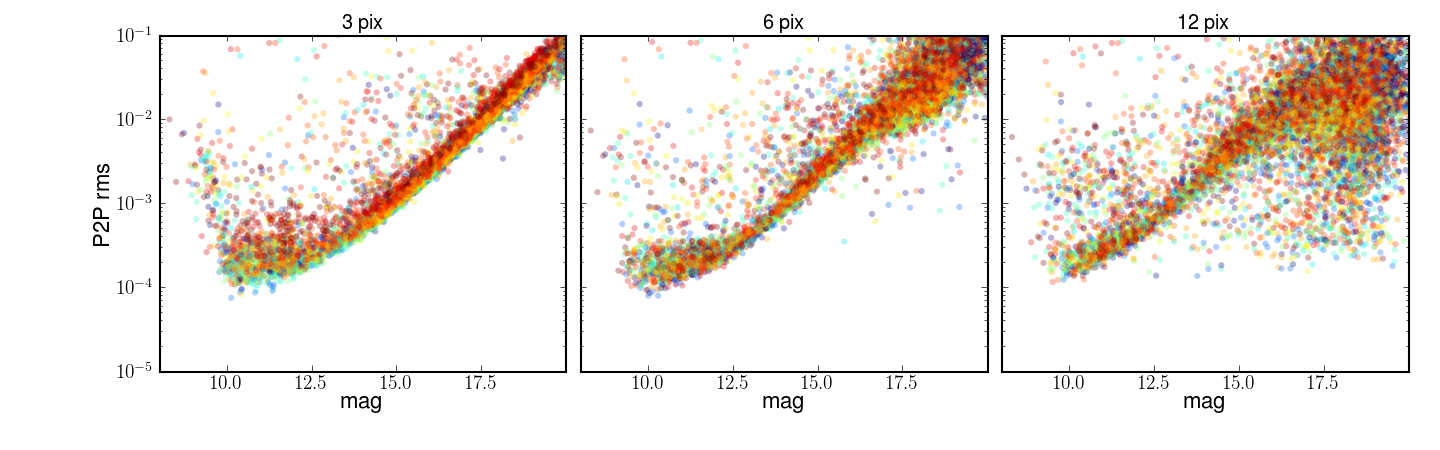}
  \caption{Point-to-point scatter as a function of magnitude using
    different aperture sizes, colour coded from blue to red according
    to increasing distance from the centre of the FOV (i.e. from the
    satellite boresight). The top panel is for the raw light curve and
    the bottom panel for the light curves after applying our
    systematics correction. See text for discussion.}
  \label{fig:p2p_bs_dist}
\end{figure*}

Using a series of fixed-size circular apertures is clearly a trade-off
between maximising the signal enclosed whilst simultenously minimising
the background and/or read-out noise included. As the \emph{Kepler}
PSF is significantly elongated at the edges of the FOV, this
necessarily implies some aperture flux loss as a function of position
in the FOV. This is illustrated in the raw light curve by the top
panel of Figure~\ref{fig:p2p_bs_dist}, which shows the point-to-point
scatter using different aperture sizes colour coded according to
boresight distance. When using small apertures, the systematics are
significantly larger at the edges of the FOV, because the aperture
collects a smaller fraction of the target flux, and the measured flux
is more sensitive to pointing variations. This is visible for the
bright stars only, because it is a small effect. The effect
essentially disappears when using larger apertures, but other issues
related to crowding become apparent.

While the variations in the detailed PSF shape over the FOV introduce
additional systematics, if these variations are predictable and
thereby repeatable, they should be correctable. Therefore, the
combination of systematics correction and magnitude-dependent aperture
selection should in principle minimize this problem. Indeed, the
effect is much smaller after systematics correction, as illustrated by
the bottom panel of Figure~\ref{fig:p2p_bs_dist}.

\subsection{Transit search}
\label{sec:trans}

\begin{table*}
  \centering
 \caption{Results of the transit search}
  \label{tab:transits}
  \begin{tabular}{rccccccccll}
    \hline
    N$^a$ & Period & Epoch$^b$ & Depth & Dur.\ & SDE & RA & DEC
    & $V$ & SIMBAD ID & Comment \\
    & (d) & (d) & (\%) & (h) & & J2000 & J2000 & (mag) & & \\ \hline 
    \multicolumn{11}{c}{Transit-like events} \\ \hline
    1981 & ---$^c$ & 2.095 & 1.14 & 2.88 & 10.37 & 00~08~57.98 &
    $+$02~56~42.0 & 10.05 & TYC 4-331-1 & known false positive \\
    11379 & 3.43249 & 3.249 & 1.36 & 2.88 & 24.84 & 23~34~27.88 &
    $-$01~34~48.1 & 11.47$^d$ & WASP-28 & known transiting planet \\ \hline
    \multicolumn{11}{c}{Detached eclipsing binaries$^e$} \\ \hline
    2550 & 1.79018 & 1.255 & 10.55 & 2.16 & 8.91 & 00~20~39.36 & $-$05~08~35.3 & 10.37 & BD-05 43 & \\  
    5041 & 3.27915 & 2.331 & 5.66 & 3.60 & 12.93 & 00~05~15.74 &
    $-$05~57~08.7 & 11.71 & TYC 4669-536-1 & \\
    5815 & ---$^c$ & 4.969 & 8.94 & 2.88 & 16.63& 23~55~37.68 &
    $-$04~22~09.9 & 12.12 & TYC 5256-1076-1 & \\ 
    5904 & 3.72184 & 2.893 & 23.85 & 3.60 & 12.32 & 00~01~47.22 &
    $-$03~10~06.5& 11.68 & TYC 4666-383-1 & \\
    7053 & 1.91282 & 2.054 & 1.81 & 1.44 & 8.84 & 23~51~02.89 &
    $-$02~35~40.8 & 11.29 & TYC 5256-76-1 & detected at P/2 \\
    10424 & 2.32830 & 1.135 & 0.43 & 1.44 & 11.21 & 23~44~58.70 &
    $-$03~36~19.7 & 11.95 & TYC 5525-818-1 & \\
    11283 & 2.95074 & 1.289 & 2.19 & 2.16 & 12.85 & 23~40~08.33 &
    $-$02~28~50.0 & 10.58 & BD-03 5686 \\ \hline
    \multicolumn{11}{c}{False positives} \\ \hline
    739 & 3.78567 & 2.230 & 0.17 & 1.44 & 8.02 & & & & & \\
    9483 & 0.61097 & 0.410 & 0.08 & 1.44 & 8.11 & & & & & \\ \hline
    \multicolumn{11}{c}{Classical variables $^f$} \\ \hline
    3998 & & & & & & & & & V* EV Psc & known RR Lyrae \\
    7214 & & & & & & & & & 0825-20053299 & contact EB \\
    9443 & & & & & & & & & 0750-21605743 & contact EB \\
    10657 & & & & & & & & & TYC 5254-832-1 & probable contact EB \\
    10847 & & & & & & & & & TYC 5255-370-1 & known contact EB \\
    11014 & & & & & & & & & NSVS 11906468 & known contact EB \\
    11553 & & & & & & & & & 0900-20441907 & contact EB \\
    12219 & & & & & & & & & V* BS Aqr & $\delta$~Scuti \\
    12271 & & & & & & & & & V* EL Aqr & known contact EB \\
    12685 & & & & & & & & & NSVS 11904371 & known contact EB \\
    13172 & & & & & & & & & NSVS 11899382 & known contact EB \\
    13336 & & & & & & & & & NSVS 11900111 & known contact EB \\
    13770 & & & & & & & & & NSVS 11899140 & known contact EB \\ \hline
  \end{tabular}

\smallskip

 Notes: $^a$: Object number in our catalog; $^b$: Epoch is given
 relative to start of run; $^c$: single transit-like even; $^d$only $I$-magnitude available for this
 object; $^e$ depth and duration for eclipsing binaries are very
 approximate as the eclipses are modelled as U-shaped: $f$ no transit
 parameters are reported for the classical variables as their true periods
 are below the minimum trial period used in the transit search.

 \end{table*}

We ran a transit search on all the objects in the engineering test
dataset with magnitude $K{\rm p} \le 14$. This was intended primarily
as a further test of the photometric pipeline, since prior experience
had shown that searching for transits can be one of the most effective
means of identifying issues in the light curves. Of course, this
exercise could also reveal previously unknown and potentially
interesting transiting planet candidates, although the relatively
small number of stars observed, and the very short time-span of the
observations, made this fairly unlikely from the outset.

The magnitude cutoff was set to ensure that any interesting candidates
identified during the search could in principle be followed up
relatively easily. A total of 1914 light curves satisfying this cutoff
were first filtered to remove long-term variations using a 1-day
iterative nonlinear filter, which consists of applying a boxcar
followed by a a median filter, flagging outliers located more than
5-$\sigma$ from the resulting smoothed light curve, repeating the
process while ignoring the outliers until no new outlyers are found,
and subtracting the final smoothed light curve from the original to
leave only the short-term variations. The iterative $\sigma$-clipping
is designed to minimize the impact of transits and eclipses on the
smoothing process.  The detrended light curves were then searched
using a standard box-shaped transit search algorithm \citep{ai04},
which is very similar to the well-known box least squares (BLS)
algorithm of \citet{kov+02}. The trial orbital period ranged from
0.5\,day to the full length of the dataset. We used a threshold of 8
in the BLS `SDE' statistic (which is obtained from the standard BLS
signal-to-noise ratio periodogram by subtracting by the mean and
dividing by the standard deviation). Table~\ref{tab:transits} lists
the 24 objects selected in this manner.

\begin{figure*}
  \centering
  \includegraphics[width=0.49\linewidth]{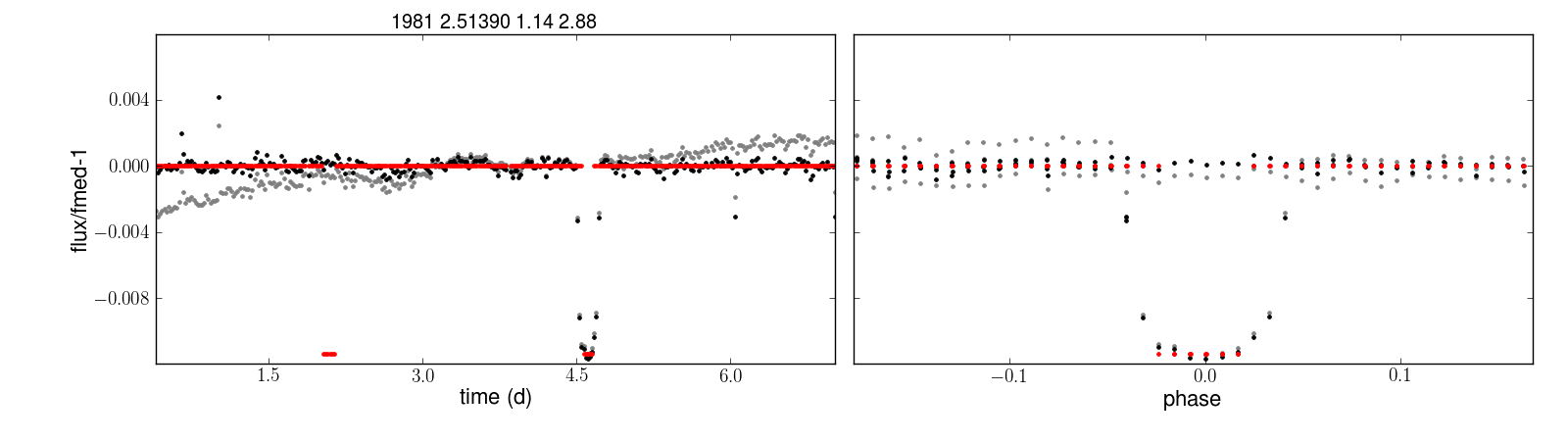}
  \includegraphics[width=0.49\linewidth]{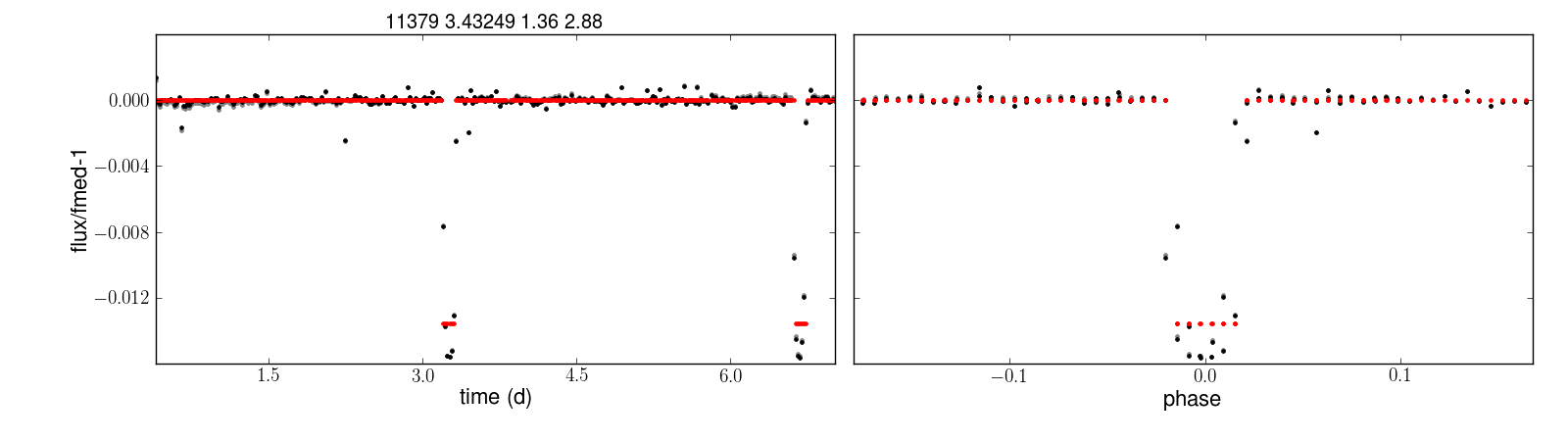}
  \includegraphics[width=0.49\linewidth]{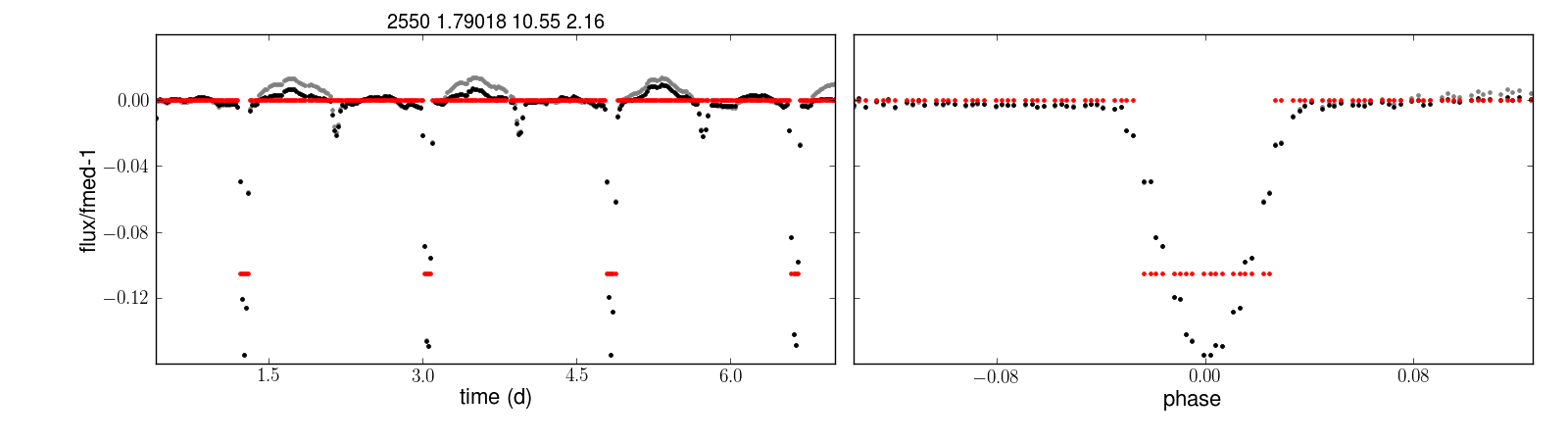}
  \includegraphics[width=0.49\linewidth]{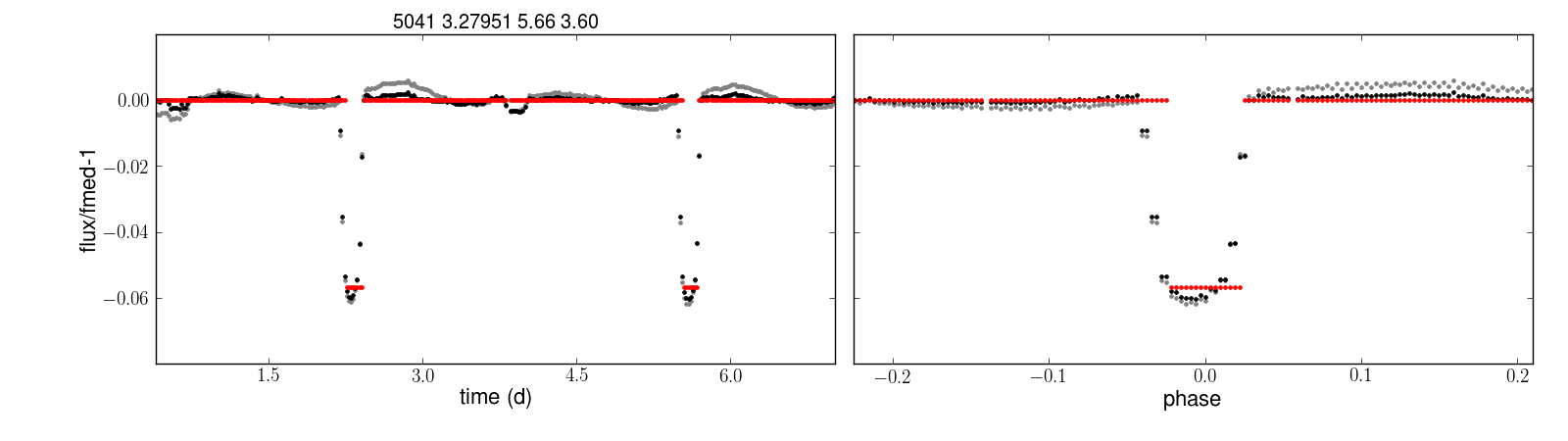}
  \includegraphics[width=0.49\linewidth]{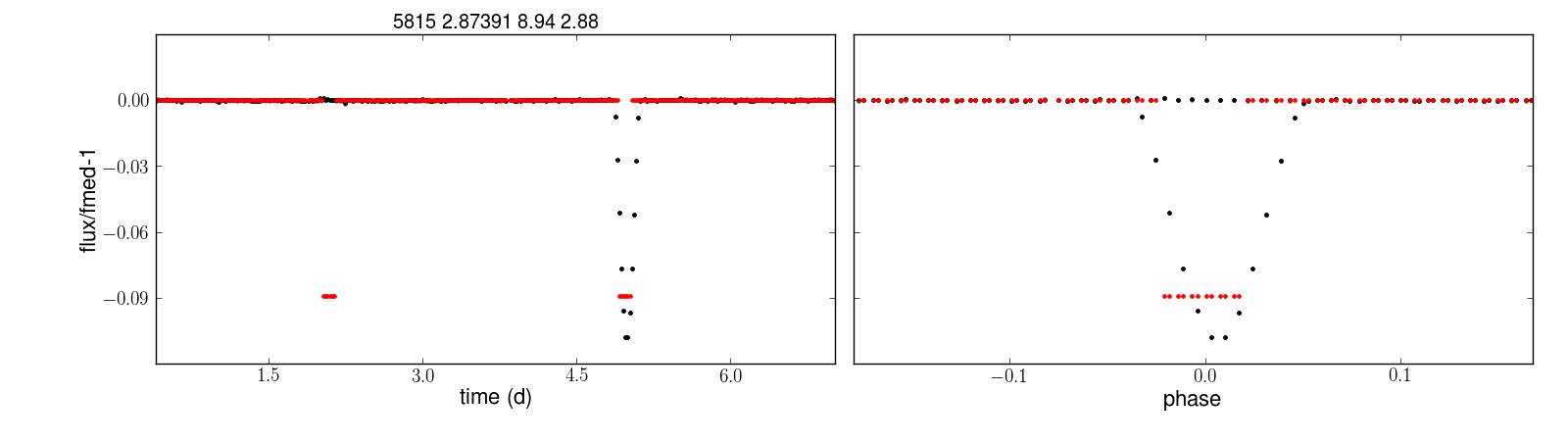}
  \includegraphics[width=0.49\linewidth]{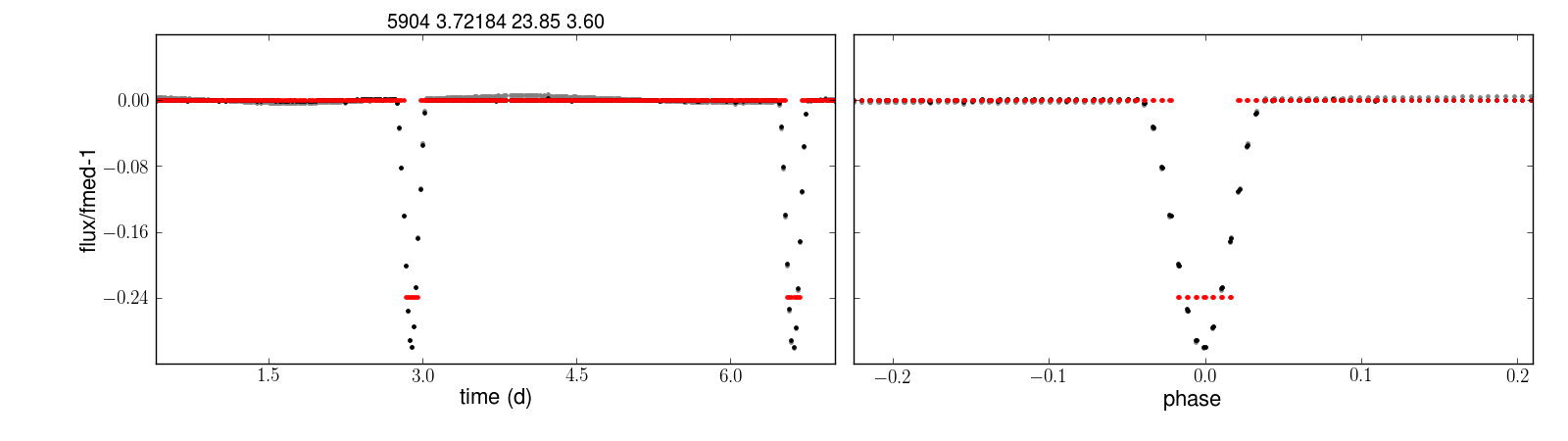}
  \includegraphics[width=0.49\linewidth]{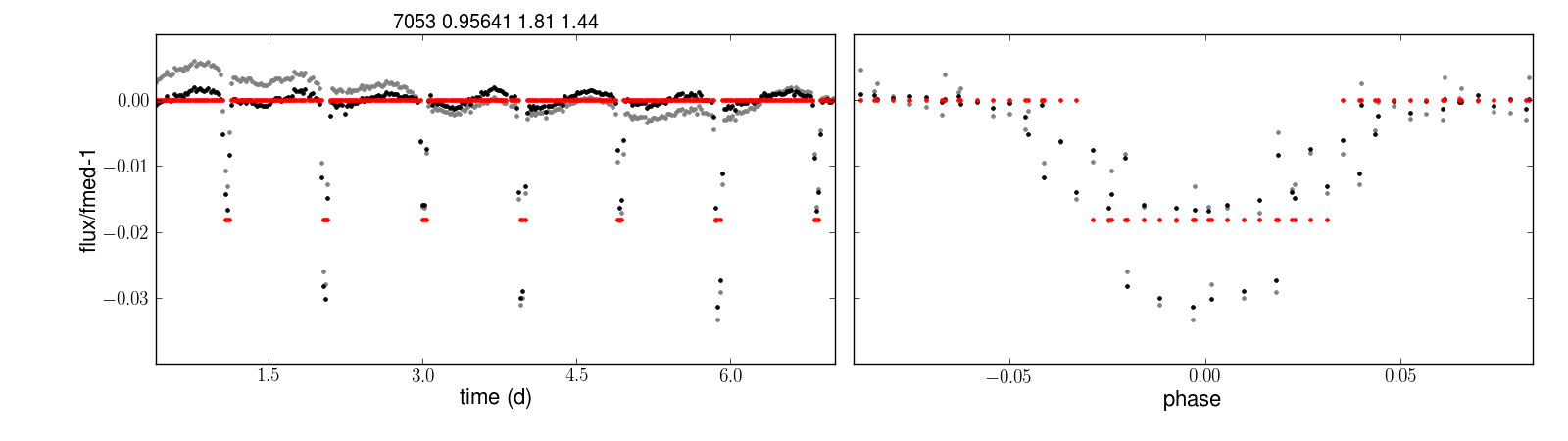}
  \includegraphics[width=0.49\linewidth]{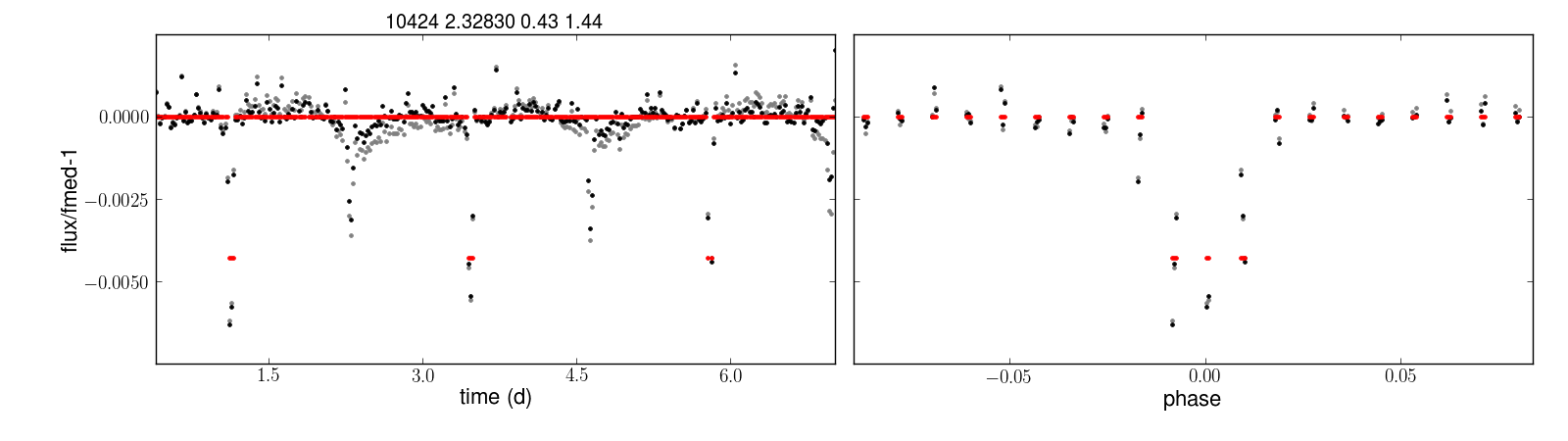}
  \includegraphics[width=0.49\linewidth]{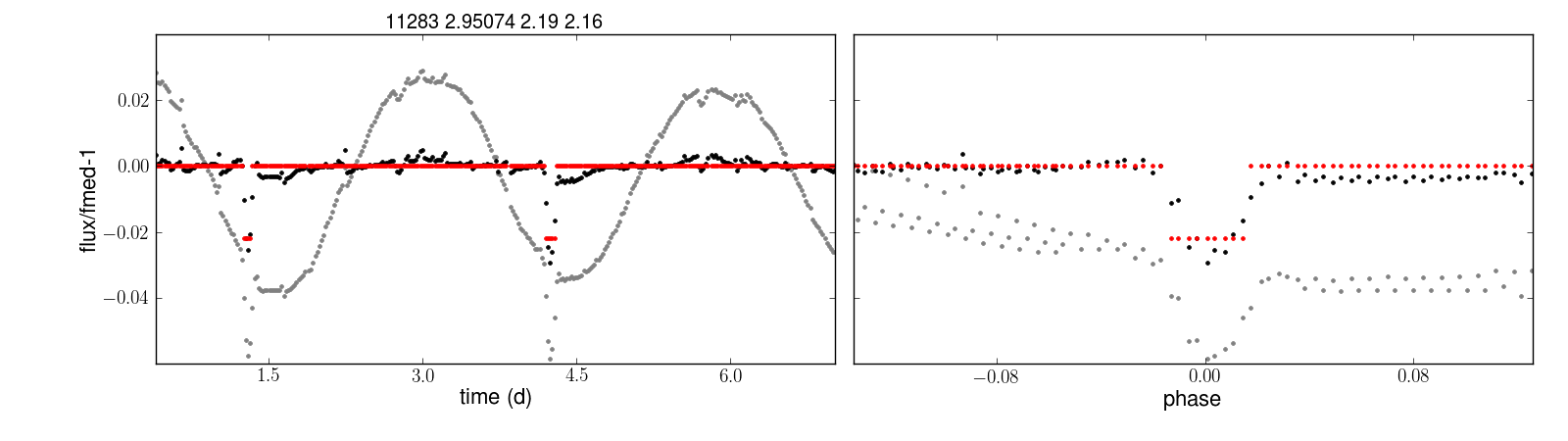}
  \caption{Transit candidates and detached eclipsing binaries
    identified in the transit search. Each pair of panels shows the unfolded and phase-folded light curves, respectively. In both
    panels, the original light curve is shown in grey, 
    the light curve after removal of long-term trends in black, and
    the best-fit transit
    model in red. }
  \label{fig:transits}
\end{figure*}

Of these, two are bona-fide transit-like events: the known transiting
planet WASP-28, and an astrophysical false positive (diluted eclipsing
binary) previously identified by the XO project \citep{pol+10}. A further seven are clearly detached
eclipsing binaries (EBs). Figure~\ref{fig:transits} illustrates the light
curves for these nine objects. The adopted threshold also selected two
objects which are clearly false positive -- `detections' caused by a
single, low outlying data point. We experimented with outlier
rejection schemes and found $5$-sigma clipping using a 3-point median
filter effectively prevents this kind of false 
positives without affecting the rest of the detections. Finally, we
detected 13 `classical' variable stars -- known RR Lyrae,
$\delta$~Scuti and contact EBs, as well as a number of
additional contact EBs for which we could not find evidence of a prior
identification in the literature. 

Overall, the results of the automated transit search are very
encouraging: the number of candidates was manageable, clearly
transit-like events were identified successfully, and the origin of
the remainder of the detections was easy to identify.

\section{Conclusions and future prospects}
\label{sec:concl}

We have presented a new method to extract high-precision light curves
from \emph{K2} data, combining list-driven aperture photometry with a
semi-parametric correction of the systematic effects associated with
the drift of the roll angle of the satellite about its boresight,
which is performed on a star-by-star basis. Finally, we propose a
simple prescription for selecting the aperture size depending on a
target's magnitude. 

We achieve a photometric precision within a factor of 2--3 of the
nominal \emph{Kepler} mission performance. From the saturation limit
($K_{\rm p} \sim 9.5$) to $K_{\rm p} \sim 12$, most stars display
scatters on transit timescales ranging from 30 to 100\,ppm, while for
fainter stars the precision is within a factor of 2 of the theoretical
noise limit. Further improvements may be achieved in the near future
by implementing a more sophisticated (spatially dependent) background
correction, and more work is needed to understand the difference
between our results and those of VJ14 for the brightest stars.  One
refinement we plan to test in the near future is to use non-circular
apertures, given the significant elongation of the PSF at the edges of
the \emph{Kepler} FOV.

A strong point of our approach is that it preserves `real'
astrophysical variabitliy, which we demonstrated using both visual
examination of individual examples and signal injection and recovery
tests. Finally, we also performed a transit search on the objects
brighter than $K{\rm p}=14$, successfully identifying the previously
known transiting planet, transit-like EB and detached EBs in the
sample.

\subsection{Implications for planet detection}

Our results and those of VJ14 have has very positive implications for
the mission's planet detection potential. They represent a factor
$\sim 2$ or better improvement over H14's early photometric precision
estimates. This confirms the expectation that \emph{K2} should readily
detect short-period gas and ice giants down to Neptune size around
bright Sun-like stars, and planets down to Earth size around
M-dwarfs. This is illustrated schematically in
Figure~\ref{fig:wasp28}, where the coloured arrows show the transit
depth that such planets would cause compared to the detection limit in
that light curve for 3 transits lasting 3 hours each.

\subsection{Computational cost for full \emph{K2} campaigns}
 
The photometric pipeline described in the present paper takes less
than a day to run on a single Intel Xeon core, on the 9-day test
dataset. To process full-length (85-day) \emph{K2} campaigns will
require considerably more computing power: while the light curve
extraction scales linearly with the amount of data, the GP regression
involved in the systematics correction (which currently accounts for
about half the total computing time) scales as the cubic power of the
number of observation. It will thus go from about 1 to 100 seconds per
star. This can be mitigated, however, by exploiting the almost
perfectly regular time-sampling of the observations, and by using
optimized matrix inversion schemes (as implemented for example in the
{\sc george} package, \citealt{amb+14}), which can reduce the scaling
from $N^3$ to $N^2$ or even $N \log N$. If we also consider the
trivially parallelizable nature of most steps in our pipeline, we do
not anticipate computational cost to be a major issue in applying the
method presented here to future \emph{K2} datasets.

\subsection{Extension to short-cadence}

Our approach uses a global astrometric solution, which relies on the
fact that there are enough stars observe across the FOV to obtain such
a solution in a reliable way. This is true for long-cadence targets
but will not, in general, hold for the much smaller number of
short-cadence targets. Our method must therefore be modified
qualitatively before it can be applied to short-cadence targets. This
is a problem which we intend to address in the near future. In the
meantime, a `shortcut' might be to proceed as follows:
\begin{itemize}
\item use standard aperture photometry (fitting for the star's
  position on each frame) to extract the raw light curve;
\item interpolate the long-cadence roll-angle measurements to the
  sampling of short-cadence data;
\item perform the systematics correction in the same way as for the
  long-cadence data, but fixing the amplitude and length scales of the
  time and roll-angle components of the GP model to the values
  obtained with long-cadence data, to keep the computational cost
  reasonable.
\end{itemize}

\section*{Acknowledgments}

The authors wish to thank the \emph{Kepler} Science Office and the
\emph{Kepler} Science Operations Center, for making the engineering
test dataset available, sharing code to compute the approximate sky
position of each module, and kindly answering numerous queries. We are
particularly indebted to Tom Barclay for his patience with our
numerous questions, and to the referee, Jon Jenkins, whose thoughful
comments helped improve the paper. We also wish to thank David Hogg,
Andrew Vanderburg \& Tsevi Mazeh for sharing their ideas on the
processing of \emph{K2} data, and for helpful discussions.

This publication makes use of data products from the Two Micron All
Sky Survey, which is a joint project of the University of
Massachusetts and the Infrared Processing and Analysis
Center/California Institute of Technology, funded by the National
Aeronautics and Space Administration and the National Science
Foundation. This research has made use of the SIMBAD database,
operated at CDS, Strasbourg, France.

This research was supported by funding from the Leverhulme Trust
(RPG-2012-661) and the UK Science and Technology Facilities Council
(ST/K00106X/1).

\bibliographystyle{mn2e}
\bibliography{K2phot}

\bsp

\label{lastpage}

\end{document}